\documentclass[prl,twocolumn,showpacs]{revtex4}
\usepackage{graphicx}
\usepackage{amsmath}

\newcommand{\celltype}[1]{{\mathrm #1}}

\newcommand{\A}{\celltype{A}}
\newcommand{\B}{\celltype{B}}

\newcommand{\area}{{\cal A}}

\newcommand{\latin}[1]{\emph{#1}}
\newcommand{\etal}{\latin{et al.}}

\newcommand{\invivo}{\latin{in vivo}}
\newcommand{\cf}{\latin{c.\,f.}}
\newcommand{\ie}{\latin{i.\,e.}}

\newcommand{\paper}{paper}

\begin{document}

\title{Cells, cancer, and rare events: homeostatic metastability\\ 
in stochastic non-linear dynamics models of skin cell proliferation}

\author{Patrick B. Warren}
\affiliation{Unilever R\&D Port Sunlight, Bebington, 
Wirral, CH63 3JW, UK.}

\date{February 12, 2009}

\begin{abstract}
A recently proposed single progenitor cell model for skin cell
proliferation [Clayton \etal, Nature {\bf 446}, 185 (2007)] is
extended to incorporate homeostasis as a fixed point of the dynamics.
Unlimited cell proliferation in such a model can be viewed as a
paradigm for the onset of cancer.  A novel way in which this can arise is
if the homeostatic fixed point becomes metastable, so that the cell
populations can escape from the homeostatic basin of attraction by a
large but rare stochastic fluctuation.  Such an event can be viewed as
the final step in a multi-stage model of carcinogenesis.  This offers
a possible explanation for the peculiar epidemiology of lung cancer in
ex-smokers.
\end{abstract}

\pacs{87.17.Ee, 87.18.Tt, 87.19.xj}

%% 87.17.Ee 	Growth and division
%% 87.18.Tt 	Noise in biological systems 
%% 87.19.xj 	Cancer

\maketitle

The epidermis is the outermost part of the skin barrier.  It comprises
10--20 layers of skin cells, which are predominantly keratinocytes
\cite{X::Gray}.  Keratinocytes proliferate in the basal layer, move up
through the middle layers, and are finally shed from the outermost
layer at a desquamation rate of the order $10^3 \,
\mathrm{cells}\,\mathrm{hr}^{-1}\, \mathrm{mm}^{-2}$ \cite{WMR84}.
This process means that cells have to proliferate continuously in the
basal layer to replenish the supra-basal layers.  Faulty cell
proliferation in the basal layer has important health-care
consequences, for example basal cell carcinoma is one of the most
prevalent forms of cancer \cite{WSL03}.  

Motivated by elegant \invivo\ experiments on labelled keratinocyte
clones, Clayton \etal\ \cite{X::ECDPD07, X::AMK07, X::AMK08} recently
proposed a novel single progenitor cell (SPC) model for basal layer
proliferation.  In the present \paper\ I examine an extension of this
SPC model to include autoregulation, and homeostasis as a dynamical
fixed point.  Interestingly, something resembling the onset of cancer
(\ie\ carcinogenesis) arises naturally in the new model if the
homeostatic fixed point loses stability in the direction of unlimited
growth of the cell populations.  One intriguing possibility is
`homeostatic metastability', in which a large but rare stochastic
fluctuation results in the cell populations escaping from the
homeostatic basin of attraction.  Such a rare escape event can be
viewed as the final step in a multi-stage model of carcinogenesis
\cite{AD54+Arm85, Frank}.

In the SPC model, there are two cell types, or `compartments':
progenitor cells A, and post-mitotic cells B.  These
proliferate according to
\begin{equation}
\begin{array}{llll}
\A\to\A+\A & \hbox{rate $\lambda_1$},\quad
&\A\to\A+\B & \hbox{rate $\lambda_2$},\\[3pt]
\A\to\B+\B & \hbox{rate $\lambda_3$},
&\B\to\emptyset & \hbox{rate $\Gamma$}.
\end{array}\label{eq:spc}
\end{equation}
The first three processes represent possible progenitor cell division
pathways.  The last process represents post-mitotic cells leaving the
basal layer.  Obviously, the progenitor cell division pathways must be
finely balanced otherwise the cell populations either grow or vanish
exponentially \cite{balance-note}.  Making the assumption therefore
that $\lambda_1 = \lambda_3$, Clayton \etal\ \cite{X::ECDPD07,
  X::AMK07} write $\lambda_1 = \lambda_3 = \lambda r$ and $\lambda_2 =
\lambda(1-2r)$.  For mouse skin, they find
$\lambda\approx0.16\,\mathrm{day}^{-1}$ for the overall progenitor
cell division rate, and $2r\approx0.16$ for the branching ratio into
the symmetric division pathways (in later work $2r\approx0.4$
\cite{X::AMK08} but the actual value is not too important).  A steady
state condition for the cell populations is
\begin{equation}
\lambda\rho = \Gamma(1-\rho)\label{eq:rho}
\end{equation}
where $\rho=n_A/n\approx0.22$ is the fraction of progenitor cells,
$n=n_A+n_B$ is the total cell density, and $n_A$ and $n_B$ are the
individual cell densities.  From this one determines
$\Gamma\approx0.045\,\mathrm{day}^{-1}$, compatible with the
above-mentioned desquamation rate.

\begin{figure}
\begin{center}
\includegraphics{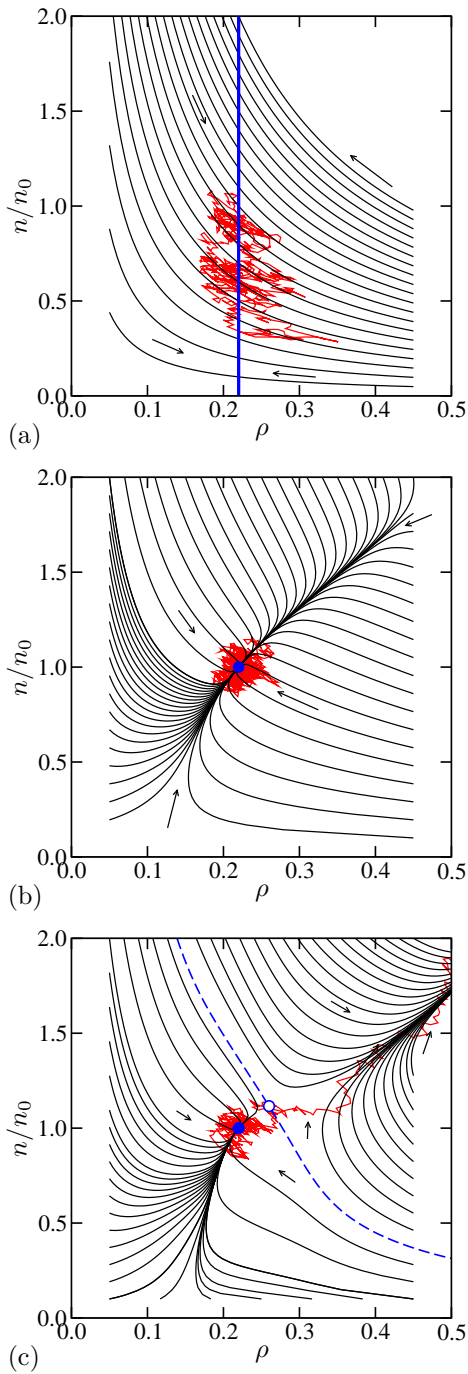}
\end{center}
\caption[?]{(color online). Phase space portraits for (a) the SPC
  model, (b) an autoregulating SPC (ASPC) model, and (c) an ASPC model
  exhibiting homeostatic metastability.  The axes are the progenitor
  cell fraction, $\rho$, and the ratio between the current and initial
  total cell densities, $n/n_0$.  Thin black lines are phase space
  flows, with the direction indicated by the arrows.  In (a) the thick
  blue line is the line of fixed points of the SPC model. In (b) and
  (c) the filled blue circles are homeostatic stable fixed points
  (nodes).  In (c) the open blue circle is an unstable fixed point (a
  saddle), lying on the homeostatic basin boundary shown as a dashed
  blue line.  Jagged red lines are representative stochastic
  trajectories.\label{fig:phase}}
\end{figure}

A phase space portrait and a representative stochastic trajectory for
the SPC model are shown in Fig.~\ref{fig:phase}(a).  The trajectory
was generated using a standard kinetic Monte-Carlo algorithm
\cite{Gill77}, starting from a patch of area $\area$ comprising
initially $N = \area n = 200$ cells, and assuming the cell types
remain well-mixed.  The actual value of $\area$ is irrelevant since
the processes in Eq.~\eqref{eq:spc} are all first order.  In
Fig.~\ref{fig:phase}, the ordinate is $n/n_0 \equiv N/N_0$ where $n_0$
($N_0$) is the initial cell density (number).

Whilst providing an eminently satisfactory explanation for the
keratinocyte labelling data, the original SPC model includes a
fine-tuning assumption, $\lambda_1=\lambda_3$, making it structurally
unstable in the language of dynamical systems theory \cite{Strogatz}.
As a consequence the model lacks homeostasis in the sense that it
possesses a \emph{line} of stable fixed points, shown in
Fig.~\ref{fig:phase}(a).  Also it is not generally robust against
perturbations, for example it cannot accommodate the introduction of a
small population of stem cells \cite{PL90}, without making some
additional fine-tuning assumptions.  One obvious solution to this is
to suppose that the cell division rates depend on the cell population
densities $n_A$ and $n_B$, representing the fact that cellular fates
are governed by integration of autocrine and paracrine signalling
factors \cite{SR85, PL90, Fuchs08}; indeed this idea was already
suggested by Jones and Simon as an avenue for further investigation
\cite{JS08}.  In such an autoregulating SPC (ASPC) model, the
fine-tuning would arise as a consequence of the cell population
dynamics driving the system to a homeostatic fixed point.  An ASPC
model would be structurally stable from the point of view of dynamical
systems theory, and also able to withstand perturbations, such as the
presence of a small number of stem cells.

To develop such an ASPC model, I start by introducing a control
parameter $q$ to describe a possible bias in the symmetric cell
division fates, thus
\begin{equation}
\lambda_1=\lambda r(1-q),\quad
\lambda_2=\lambda(1-2r),\quad
\lambda_3=\lambda r(1+q).
\end{equation}
The parameters $\lambda$, $r$ and $q$ replace $\lambda_1$, $\lambda_2$
and $\lambda_3$ without loss of generality.  The steady state
conditions are given by $q=0$ and Eq.~\eqref{eq:rho}.  From this point
onwards, I deliberately adopt a `physics-oriented' approach in which
the model is kept as simple as possible to expose the general
mechanisms at work.  In particular it is not claimed that the biology
is accurately represented.  The basic idea is to introduce a
minimalist dependence of $q$ and $\lambda$ on the cell population
densities, to represent the integrated effect of the intercellular
signals.  I start by making $q=q(\rho)$.  Then the steady state
condition selects the value of $\rho$ for which $q(\rho) = 0$.  I
additionally suppose that $\lambda=\lambda(n)$ is a decreasing
function of the total number density ($\lambda'<0$) representing the
fact that the progenitor cell proliferation rate should be reduced
if the overall cell density increases \cite{gamma-note}.  With
these assumptions, fixed points of the dynamics are determined by
$q(\rho)=0$ and $\lambda(n)=\Gamma(1-\rho)/\rho$.  It is a
straightforward exercise to show, in the language of dynamical systems
theory, that a fixed point is a stable node if $q'>0$, and a saddle if
$q'<0$.

A concrete model of this type has 
\begin{equation}
\lambda=\lambda_0
\Bigl(\frac{n_0}{n}\Bigr)^\alpha,\quad
q=\tanh\Bigl[
\frac{\beta\rho_0(1-\rho_0)(\rho-\rho_0)}{\rho(1-\rho)}
\Bigr],\label{eq:lq}
\end{equation}
where $\lambda_0 = {\Gamma(1-\rho_0)}/{\rho_0}$.  This model has a
stable node at a target population density $n=n_0$ and progenitor cell
fraction $\rho=\rho_0$.  I emphasise that \emph{these functions have
  been arbitrarily chosen for illustrative purposes}, though with care
to make sure that they have the appropriate limiting behaviours. In
Eq.~\eqref{eq:lq} $\alpha$ and $\beta$ are `stiffness'
coefficients. They are related to derivatives of the functions at the
fixed point by $\alpha=-{n \lambda'}/{\lambda}$ and $\beta=q'$.  For
results presented below I use $\alpha=2$ and $\beta=10$ which allows
for stochastic fluctuations of moderate amplitude, balanced between the
two cell types.  I have checked that the results are qualitatively
insensitive to this choice.

Fig.~\ref{fig:phase}(b) shows the phase space portrait and a
representative stochastic trajectory for this ASPC model with a fixed
point chosen to lie at $\rho_0=0.22$ and a target population size
$N_0=\area n_0=200$.  Again $\area$ does not need to be explicitly
specified though in this case it could be interpreted as representing
the area of influence of diffusible intercellular signalling factors.
In contrast to Fig.~\ref{fig:phase}(a), there is an
isolated stable fixed point, whose basin of attraction extends to
cover the whole plane.  The stochastic trajectory is strongly
localised to the vicinity of the fixed point.  It is clear that this
behaviour should be generic to a wide class of ASPC models since the
existence of an isolated stable fixed point is a structurally stable
feature of the dynamics.  The fixed point in a model of this type
represents homeostasis in several ways.  Firstly, if the cell
populations are perturbed, they will tend to return to the original
fixed point.  Secondly, fluctuations in the cell populations will be
limited to the vicinity of the fixed point.  Thirdly, the model itself
can be perturbed, for example by the inclusion of stem cells, without
leading to a qualitative change in behaviour.

A key requirement of such models is that they retain compatibility
with the keratinocyte labelling data of Clayton
\etal\ \cite{X::ECDPD07, X::AMK07}.  An (admittedly mean-field)
argument that this is true can be made as follows.  Imagine labelling
a \emph{small} fraction of keratinocytes.  If the label is
\emph{passive}, proliferation of labelled cells will be determined by
\emph{fixed} parameter values, corresponding to the homeostatic fixed
point.  In particular the symmetric division pathways of labelled
progenitor cells will be automatically balanced.

I now turn to the implications for cancer modelling.  As outlined in
the introduction, the ASPC models might offer some interesting
insights into carcinogenesis, which can be considered as unlimited
cell proliferation caused either by a loss of stability of the
homeostatic fixed point, or by a large but rare stochastic fluctuation
causing the system to exit the homeostatic basin of attraction.
Whilst the idea that cancer arises from an instability in the
underlying cell population dynamics is rather old \cite{Whel73-BMVP96,
  SR85}, the second possibility is very intriguing and has apparently
not been hitherto considered.  My studies of models comprising $\A$,
$\A^*$ and $B$ cells, with a process $\A\to\A^*$ representing a
somatic mutation (\cf\ Klein \etal\ \cite{X::AMK07}), shows that the
phenomenon of `homeostatic metastability' can easily be observed.
However the resulting three-dimensional phase space portraits are
tricky to represent and analyse.  To illustrate the mechanism of
homeostatic metastability therefore, I return to the original ASPC
model, but introduce a \emph{re-entrant} bias control function $q(\rho)$.
Such a model is a prototypical example of a system which is near a
saddle-node bifurcation.

An example of a re-entrant $q(\rho)$ is given by inserting an extra
factor $(\rho_1-\rho)/(\rho_1-\rho_0)$ in the argument to the tanh
function in Eq.~\eqref{eq:lq}.  This model has a stable node at
$\rho = \rho_0$ and a saddle at $\rho = \rho_1$.  The saddle-node
bifurcation is approached as $\Delta\rho = \rho_1 - \rho_0$ vanishes.
The phase space portrait and a representative stochastic trajectory
for this ASPC model are shown in Fig.~\ref{fig:phase}(c) for
$\Delta\rho=0.04$ (other parameters as for the original ASPC model).
The homeostatic basin of attraction is now confined to the lower left
region.  The saddle lies on the homeostatic basin boundary.  The
simulations show that the system inevitably escapes from the
homeostatic basin, typically in the vicinity of the saddle. After this
the cell populations grow without limit.

\begin{figure}
\begin{center}
\includegraphics{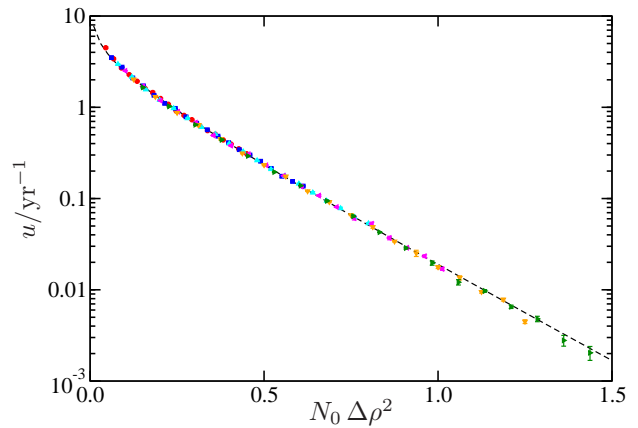}
\end{center}
\caption[?]{(color online).  Data collapse of homeostatic escape rate
  $u$ as a function of $N_0\Delta\rho^2$.  Data were collected for
  various $N_0$ and (colored) for
  $\Delta\rho=0.030(0.005)0.055$.\label{fig:escape}}
\end{figure}

To characterise the escape event, I generate a large set of
trajectories and compute an escape rate $u$ from the probability of
remaining in the homeostatic basin \cite{mfpt-note}.
Fig.~\ref{fig:escape} shows that $u$ decreases approximately
exponentially with $N_0\Delta\rho^2$ \cite{dep-note}.  The sensitive
dependence on the target population size $N_0$ is reminiscent of
system size effects found for other non-equilibrium phase transitions
\cite{MREWtW}.  The sensitivity to the distance $\Delta\rho$ from the
saddle-node bifurcation may reflect a quadratic dependence of the
height of an `action' barrier \cite{AS02}.

The concept of homeostatic metastability fits neatly into multi-stage
models which are widely used to interpret cancer epidemiology
\cite{AD54+Arm85, Frank}.  In a multi-stage model, cell lineages
slowly transit through one or more pre-cancerous stages before
entering a final cancerous stage.  The idea presented here is that
homeostatic escape could be the final slow step after one or more
somatic mutations have taken place, bringing, for
instance, the system close to a saddle-node bifurcation.  Note that,
extrapolating from Fig.~\ref{fig:escape}, the homeostatic escape rate
can easily be comparable to multi-stage transition rates which are of
the order $10^{-6}\text{--}10^{-4}\,\mathrm{yr}^{-1}$ \cite{Frank04b}.
In this context it is intriguing to note that the initial step in the
development of skin cancer often appears to involve a mutation in the
ras signalling pathway \cite{Barb87}, which is important for
controlling cell proliferation.

The mechanism of epithelial renewal in the lungs has much in common
with that of the skin, although the turnover time may be somewhat
longer \cite{RH06}.  Lung cancer might therefore be expected to have
many features in common with skin cancer.  A long-standing puzzle in
multi-stage models of lung cancer has been the apparent indifference
of the rate of the final step to the presence of a carcinogen (tobacco
smoke) \cite{HGW93-Cairns02, Cairns06, Peto01}.  This
epidemiology has been interpreted as indicating that a non-mutagenic
mechanism might be at work.  One suggestion is that the final step is
epigenetic in nature \cite{Cairns06}.  Another suggestion is that the
final step somehow involves signalling \cite{Peto01}.  Homeostatic
metastability can be seen as a specific example of the latter
signalling mechanism.  As Frank challenges though \cite{Frank}, any
such interpretation of the epidemiology should be supported by other
evidence.  Experiments that directly test the mechanism of homeostatic
escape are therefore very desirable!  A central feature of the
present model, around which experiments may perhaps be designed, is
that the cells themselves do not undergo any change if the system
escapes from homeostasis; only the micro-environment changes.

Some general points can be made about directions for future research.
Firstly, many of the arguments presented here are mean-field in
nature.  This key point was well appreciated by Klein
\etal\ \cite{X::AMK08}, who showed that many mean-field results still
hold in a two-dimensional version of the original SPC model.  The
extension of the present ASPC models to fully-fledged two dimensional
models is clearly a crucial next step.  Another direction in which
progress could be made is to improve the representation of the
biology, for example moving to multi-scale \cite{AQ08}, or agent-based
models \cite{X::Sun+07}, which capture the detailed biology of the
intercellular signals, and also the essential stochastic nature of
individual cell fates.

A generic conclusion of the present study is that it may not be valid
to examine just the \emph{deterministic} consequences of somatic
mutations, since rare stochastic events may occur at comparable rates.
This makes the task of examining the behaviour of more biologically
detailed models rather formidable.  Brute force methods (\ie\ lots of
very long simulations) have been used in the present \paper\ since the
underlying stochastic processes are rather simple.  This may not be
possible for more complex models.  Instead, it may be necessary to
bring to bear more sophisticated techniques such as transition path
sampling \cite{X::DBCC98}, or forward-flux sampling \cite{AWtW05}.

I thank Rosalind Allen, Mike Cates and Martin Evans for helpful
discussions and encouragement.

%\bibliography{skin,aspc4p3x}

\begin{thebibliography}{31}
\bibitem[{bal()}]{balance-note}
\bibinfo{note}{One might require $T\delta\lambda\alt 1$ where $\delta\lambda =
  |\lambda_1-\lambda_3|$ and $T$ is the lifetime for the organism: for a mouse
  this implies $\delta\lambda/(\lambda r)\alt0.2$; for a human
  $\delta\lambda/(\lambda r)\alt4\times10^{-3}$.}
\bibitem[{gam()}]{gamma-note}
\bibinfo{note}{Very similar results are obtained if instead one makes
  $\Gamma=\Gamma(n)$ an increasing function of $n$, corresponding to an
  increasing exit rate of post-mitotic cells.}
\bibitem[{mfp()}]{mfpt-note}
\bibinfo{note}{The method is essentially equivalent to computing the first
  passage time distribution.}
\bibitem[{dep()}]{dep-note}
\bibinfo{note}{The dependence of $u$ on $\alpha$ and $\beta$ is more complex
  and not necessarily monotonic.}
\end{thebibliography}

\end{document}